\documentclass{elsart1p}
\usepackage{graphics}
\usepackage{epsfig}
\usepackage{graphicx}
\usepackage{amssymb}
\begin{document}
\begin{frontmatter}
%
%
%
%
%
\title{$\rho^{0}$ - vector meson elliptic flow ($v_{2}$) in $Au+Au$
  collisions at $\sqrt{s_{NN}} = 200$ GeV in STAR at RHIC}
%
%

\author{Prabhat R. Pujahari (for the STAR collaboration)}

\address{Department of Physics, Indian Institute of Technology bombay,
  Mumbai - India}

\begin{abstract}
The first measurement of the $\rho^{0}$ - vector meson
elliptic flow $v_{2}$ at mid-rapidity ($|y |$ $<$
0.5) in $40 - 80$ $\%$ centrality in $Au+Au$ collisions
at $\sqrt{s_{NN}} = 200$ GeV from the STAR experiment at RHIC is presented. The
study is through the $\pi^{+} \pi^{-}$ hadronic decay channel of
$\rho^{0}$ which has a branching ratio of $\sim$ 100 \%. The analysis is being carried out in two different
methods. The $v_{2}$ results obtained in these methods are
consistent. Number of Constituent Quark (NCQ) scaling of $v_{2}$ of
$\rho^{0}$ meson with respect to other hadrons at intermediate $p_{T}$
is observed. The $\rho^{0}$ $v_{2}$ favors $NCQ = 2$ scaling, supporting
the coalescence being the dominant mechanism of hadronization in the
intermediate $p_{T}$ region at RHIC.

\end{abstract}

%
%

\end{frontmatter}

\section{Introduction}
\label{}
The primary aim of ultra-relativistic heavy-ion collisions is to
produce and study a state of high-density nuclear matter called the
Quark-Gluon Plasma (QGP). In the search of this new form of matter,
penetrating probes are essential in order to gain information from the
early stage of the collisions. The lifetime of the $\rho^{0}$ meson
is about 1.3 fm/c, which is smaller than the life time of the system
formed in $Au+Au$ collisions at such energy. The $\rho^{0}$ measured via its hadronic
decay channel (branching ratio $\sim$ 100$\%$) can be used as a sensitive tool to examine the collision
dynamics in the hadronic medium through its decay and regeneration. \\

Elliptic flow, $v_{2}$, is an observable which is thought to reflect
conditions from early stage of the collisions [1, 2]. In non-central
heavy-ion collisions, the initial spatial anisotropy of the overlap
region of the colliding nuclei is transformed into an anisotropy in
momentum space through interactions among the produced particles.
Systematic measurements of the $v_{2}$ of hadrons show that the
$v_{2}$ scales with the number of constituent quarks in the
intermediate $p_{T}$ region ($1.5 \leq p_{T} \leq 5$ GeV/c). It has been
proposed that the measurement of the $v_{2}$ of resonances can
distinguish whether the resonances were produced 
from a hadronizing quark gluon plasma (QGP-mechanism) or in the
hadronic final state via hadron-hadron rescattering (HG mechanism) [3].

\section{Results}
\label{}
The results presented in this paper were obtained with the STAR
detector [4] at the Relativistic Heavy Ion Collider (RHIC) at
Brookhaven National Laboratory (BNL), USA. The sub-detectors used in this
analysis were the Time Projection Chamber (TPC), and the trigger
detectors, namely Zero Degree Calorimeter (ZDC). The collision centrality
was determined by charged hadron multiplicity measured in TPC within the
pseudo-rapidity $|\eta|$ $<$ 0.5. \\

\begin{figure}[ht]
  \begin{center}
\includegraphics[scale=0.3]{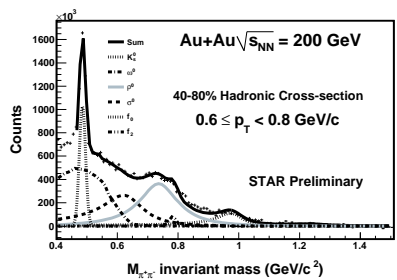}
\caption{\label{fig1} $\pi^{+}\pi^{-}$ invariant mass distribution after
background subtraction in $Au+Au$ collisions.}
\end{center}
\end{figure}

The $\rho^{0}$ yield in each $p_{T}$ bin was extracted from the
invariant mass ($m_{inv}$) distribution of $\pi^{+}$ and $\pi^{-}$
candidates after subtraction of like-sign combinatorial background 
obtained from the geometric mean of the $\pi^{+} \pi^{+}$ and $\pi^{-}
\pi^{-}$ invariant mass distributions in the same event. The $\pi^{+}  \pi^{-}$
invariant mass distribution and the combinatorial background are
normalized in the invariant mass range from 1.5 GeV/$c^2$ to 2.5 GeV/$c^2$
before subtraction. The pions were identified
through their $\emph{dE/dx}$ energy loss in the STAR TPC [4]. A
typical $\pi^{+} \pi^{-}$ invariant mass distributions after background
subtraction for $40 - 80$$\%$ centrality and $0.6 \leq p_{T} < 0.8$ GeV/c in $Au+Au$ collisions at
$\sqrt{s_{NN}} = 200$ GeV is shown in Fig. 1. The solid black line in Fig. 1 is
the sum of all the contributions in hadronic cocktail. The $K_{s}^{0}$
was fit to a gaussian. The $\omega$ shape was obtained from the HIJING
event generator [5]. The $\rho^{0} (770)$, the $f_{0} (980)$, the
$f_{2}$ (1270) and the $\sigma^{0}$ were fit by relativistic
Breit-Wigner functions [6] times the Boltzmann factor which 
accounts for the phase space [7,8] in the
hadronic cocktail. In the cocktail fit, the $\rho^{0}$ width was fixed
at 160 MeV/$c^{2}$. The $\sigma^{0}$ mass and width were fixed at
630 MeV/$c^{2}$ and 160 MeV/$c^{2}$, respectively. The temperature in the phase space was taken to be 120 MeV
[8]. \\

Two different techniques are being used to find out the $v_{2}$ of the
$\rho^{0}$-meson. One is $v_{2}$ vs. invariant mass method [9] and the
other one is ($\phi - \Psi_{2}$) method [10]. The invariant mass method involves
calculating the $v_{2}$ of the same-event distribution as a function
of $m_{inv}$ and then fitting the resulting $v_{2}(m_{inv})$
distribution using a multi parameters function: \\
\begin{equation}
v_{2} (m_{inv}) = v_{2S} \alpha (m_{inv})  + v_{2B} (m_{inv}) \beta (m_{inv})
\end{equation}
where $v_{2S}$ is the signal $v_{2}$ and $v_{2B}$ is the background
$v_{2}$. The signal $v_{2S}$ contribution is coming from $\rho^{0}$,
$\sigma^{0}$, $\omega^{0}$, $K_{S}^{0}$, $f_{0}$ and $f_{2}$, i.e. $v_{2S} = v_{2\rho^{0}} + v_{2\sigma^{0}} + v_{2\omega^{0}} +
v_{2 K_{s}^{0}} + v_{2 f_{0}} + v_{2 f_{2}} $.
The background $v_{2B}$ is calculated from the
$\pi^{+}\pi^{+}$ and $\pi^{-}\pi^{-}$ pairs $v_{2}$.
The parameters $\alpha (m_{inv}) =
S/(S+B)$ and $\beta (m_{inv}) = B/(S+B)$ where $\emph{S}$ is the sum of all the individual particles signal in
the cocktail and $B$ is the background contribution from the $\pi^{+} \pi^{+}$
and $\pi^{-} \pi^{-}$ invariant mass distributions.
Fig. 2 represents the $v_{2}$(Total) as a function of invariant mass for a particular
$p_{T}$ bin ($p_{T} = 0.7$ GeV/c). The $v_{2}$(Total) is the total $v_{2}$ of all the
$\pi^{+} \pi^{-}$ combinations in the same event. The signal $v_{2}$ is extracted by doing a
fit using Eq.(1) to the $v_{2}$(Total) as shown in Fig. 2.

\begin{figure}[ht]
\begin{minipage}[b]{0.5\linewidth}
\centering
\includegraphics[scale=0.25]{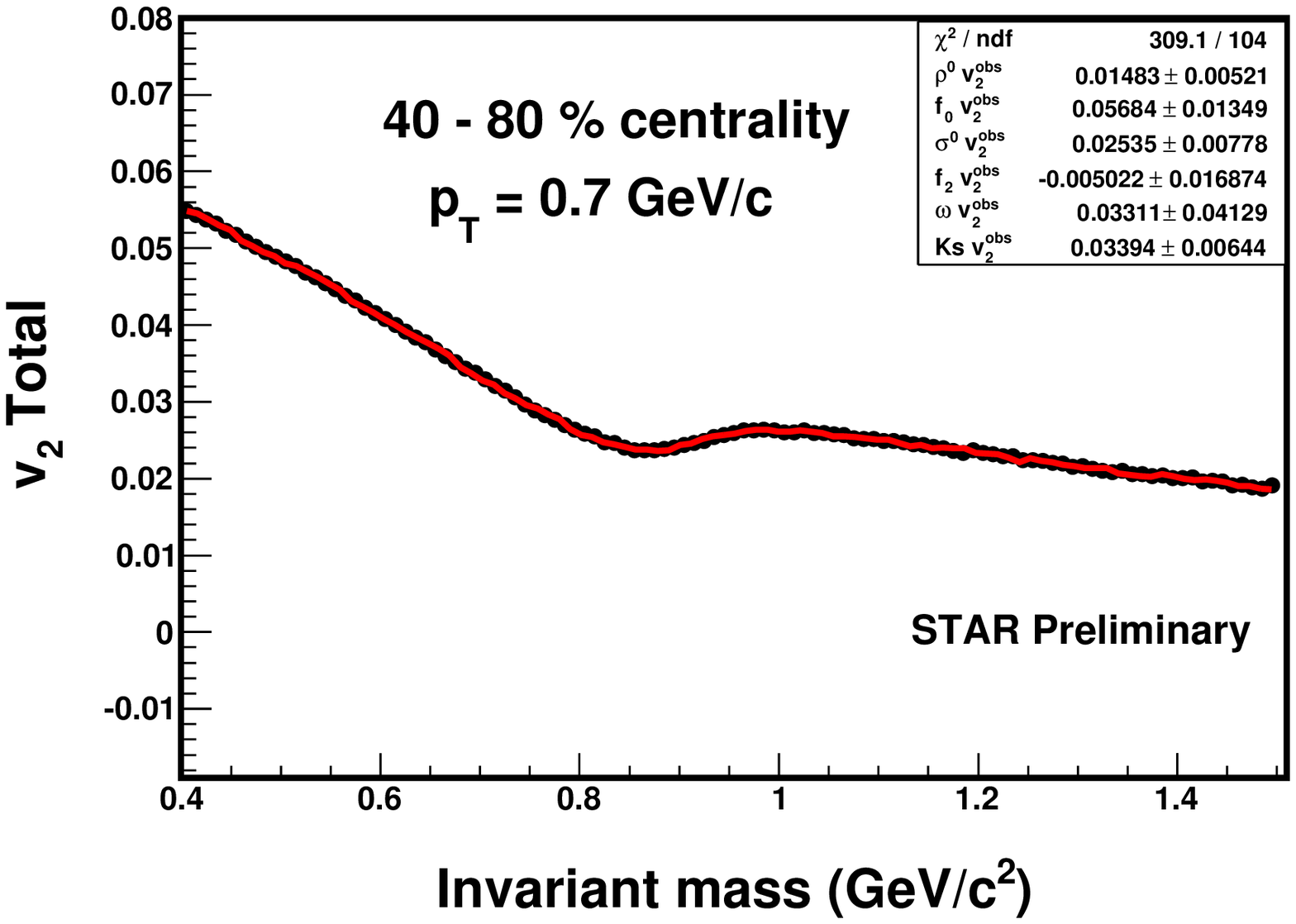}
\caption{The $v_{2}$ vs. invariant mass. The solid line is the result
  from the fitting function.}
\label{fig:figure1}
\end{minipage}
\hspace{0.5cm}
\begin{minipage}[b]{0.5\linewidth}
\centering
\includegraphics[scale=0.25]{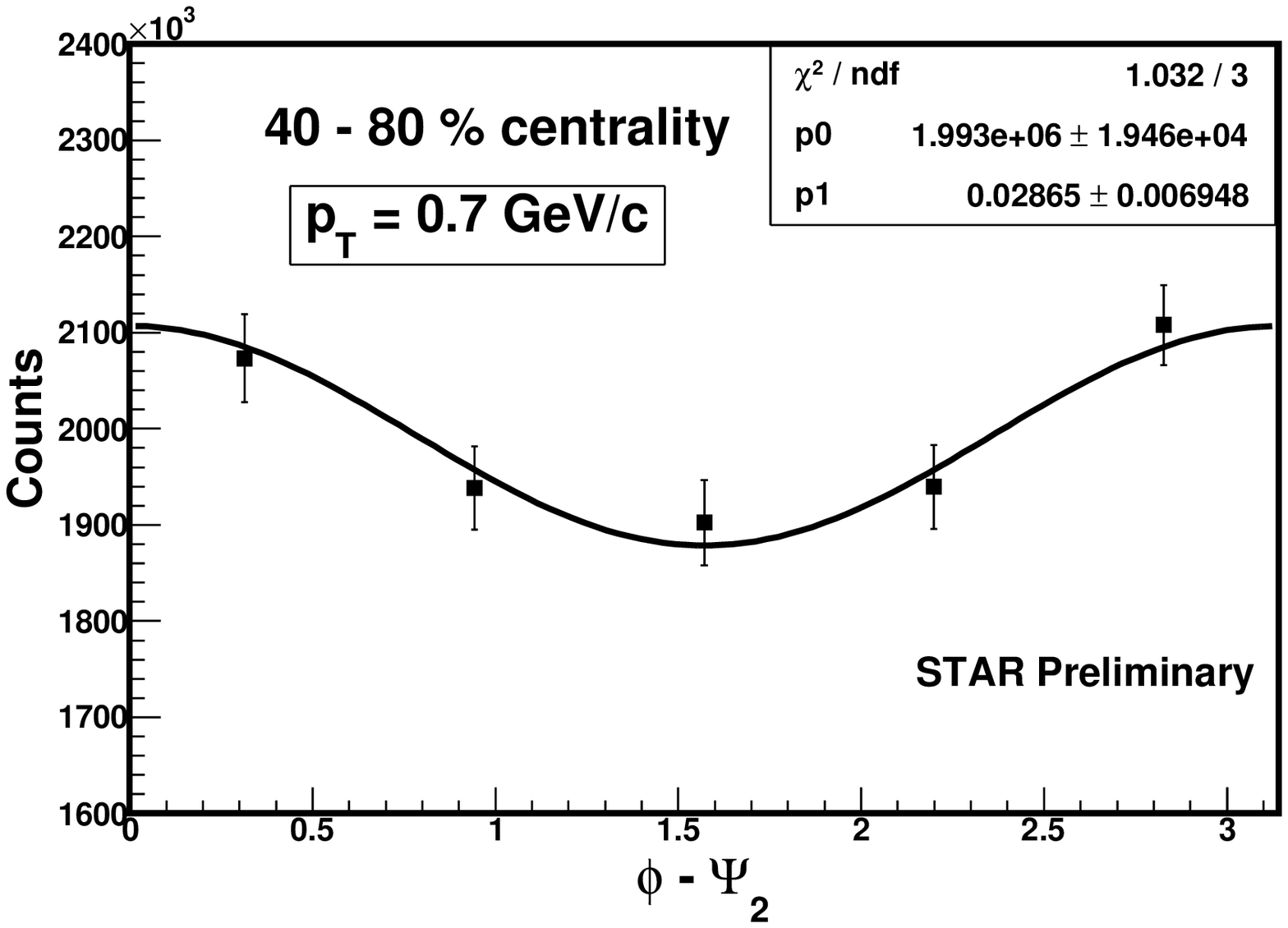}
\caption{The $\rho^{0}$ counts as a function of $\phi - \Psi_{2}$. The
solid line is the result of the fitting function.}
\label{fig:figure2}
\end{minipage}
\end{figure}

In order to compare the results obtained in the above method, we used
another method to calculate the $v_{2}$ of $\rho^{0}$ which is called
the standard ($\phi - \Psi_{2}$) method and described in the reference
[10].
Fig. 3 shows the $\rho^{0}$-meson yield after background subtraction as a function
of ($\phi - \Psi_{2}$). The observed $v_{2}$ is obtained from the distribution by fitting a function of
the form  $dN/d\phi = P_{0}[1 + 2v_{2} \cos(2(\phi -\Psi_{2}))]$. The observed $v_{2}$
parameters were corrected for the 
event plane resolution to get the final $v_{2}$
values. Fig. 4 shows the corrected $v_{2}$ as a function of
$p_{T}$. Solid closed circles are the data points obtained from the invariant mass technique
whereas the open closed circles are for the standard ($\phi -
\Psi_{2}$ bin) method. It is
clear from the figure that the results obtained in both the techniques
are consistent within the statistical error. Fig. 5 represents the comparison of $\rho^{0}$ $v_{2}$ with $K_{s}^{0}$ and $\Lambda^{0}$
$v_{2}$ for the same centrality class, i.e. $40 - 80$$\%$, in $Au+Au$ collisions.
The solid circles are the data points for $\rho^{0}$ $v_{2}$ obtained
from invariant mass method. The open
circles are the data points for $K_{s}^{0}$ and open squares are for
$\Lambda^{0}$. The $K_{s}^{0}$ and $\Lambda^{0}$ data points are taken from [11].
It is clear from Fig. 5 that the $v_{2}$ of
$\rho^{0}$ is more close to the $v_{2}$ of $K_{s}^{0}$ than
$v_{2}$ of $\Lambda^{0}$ in the region $p_{T}$ $>$ 1.5 GeV/c.
In the low $p_{T}$ region ($\sim 0.3-1.0$
GeV/c) $\rho^{0}$-meson seems to deviate from the usual mass
ordering. \\

\begin{figure}[ht]
  \begin{center}
\includegraphics[scale=0.27]{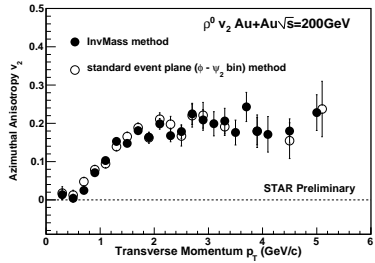}
\caption{\label{fig4} Comparison of $\rho^{0}$ $v_{2}$ obtained in
  $v_{2}$ vs. invariant mass method and the standard event plane
  method. Only the statistical error bars were shown in the plot.}
\end{center}
\end{figure}
The number of constituent quark (NCQ) scaling of $v_{2}$ of
$\rho^{0}$-vector meson in the intermidiate $p_{T}$
 is shown in Fig. 6. The $K_{S}^{0}$ and $\Lambda$ $v_{2}$ are
 plotted in the same figure for comparison after scaling with $n=2$ and $n=3$ quarks,
respectively. This measurement shows that the $\rho^{0}$ $v_{2}$
scales with $n=2$ quarks in the intermediate $p_{T}$ range.

\begin{figure}[ht]
\begin{minipage}[b]{0.5\linewidth}
\centering
\includegraphics[scale=0.27]{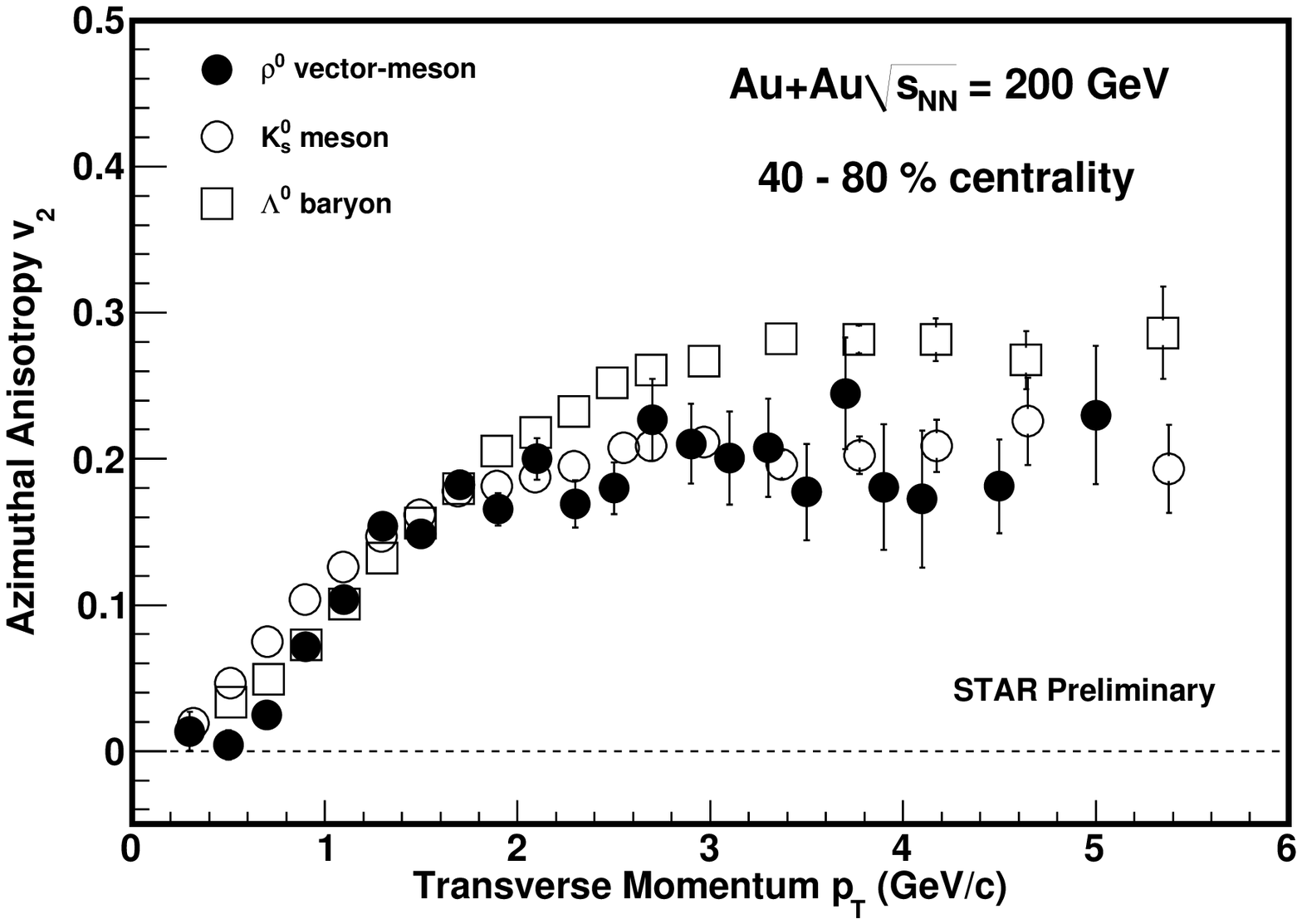}
\caption{\label{fig2} $p_{T}$ dependence of elliptic flow ($v_{2}$) of
  $\rho^{0}$-meson in $Au+Au$ collisions ($40 - 80$ $\%$
  centrality). Data poins of $\rho^{0}$ meson are from $v_{2}$
  vs. invariant mass method. The vertical error bars represent the statistical errors.}
\label{fig:figure1}
\end{minipage}
\hspace{0.5cm}
\begin{minipage}[b]{0.5\linewidth}
\centering
\includegraphics[scale=0.27]{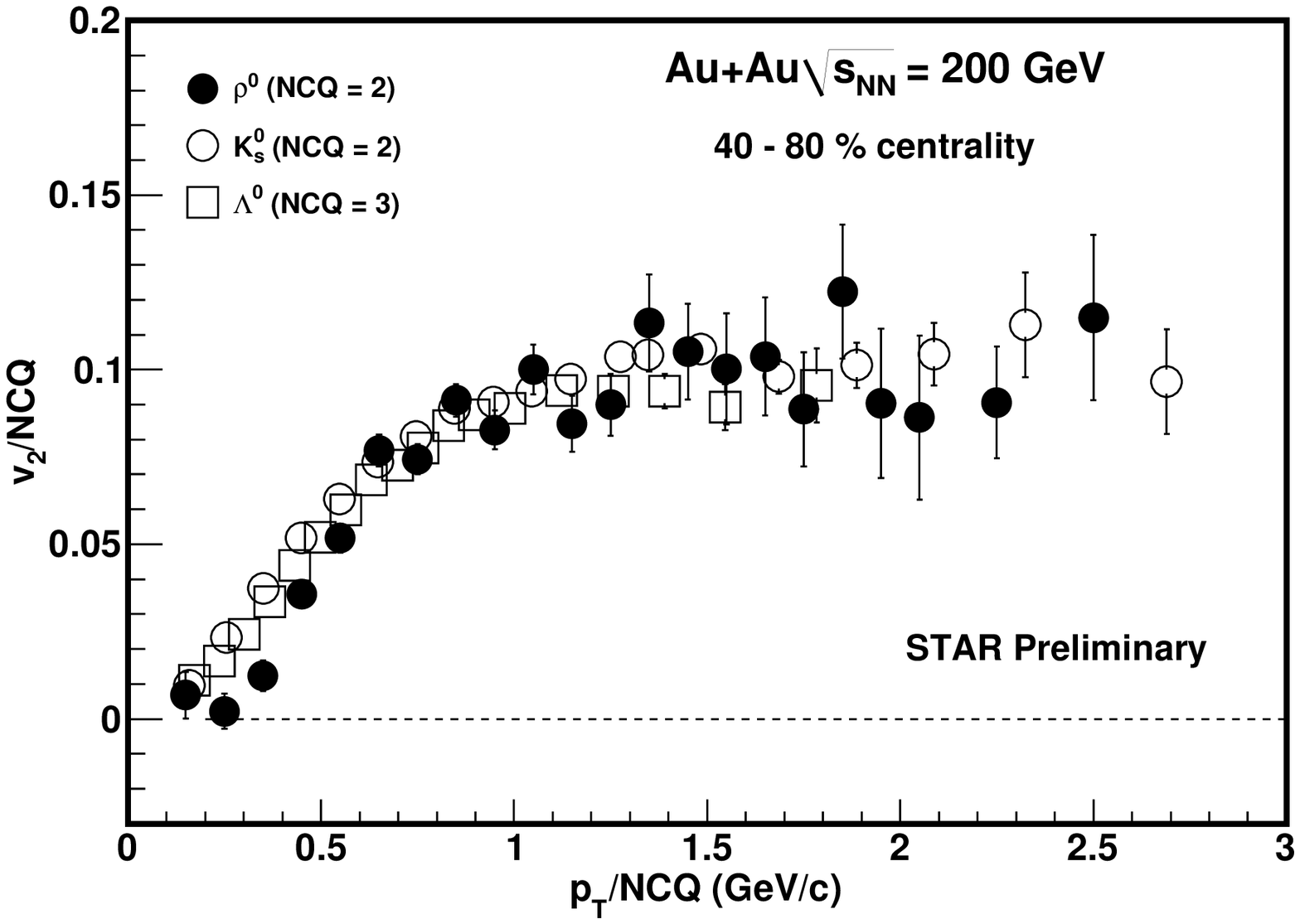}
\caption{\label{fig2} Constituent quark number scaling of elliptic
  flow ($v_{2}$) for the $\rho^{0}$-meson, the $K_{s}^{0}$ meson and
  the $\Lambda^{0}$ baryon in $Au+Au$ $200$ GeV for $40 - 80$$\%$ centrality. The vertical error
     bars represent the  statistical errors.}
\label{fig:figure2}
\end{minipage}
\end{figure}

\section{Conclusion}
\label{}
The $\rho^{0}$ $v_{2}$ is measured in $Au+Au$ collisions at 200 GeV  in two
different methods and the results obtained are consistent within the
statistical errors. From the
number of constituent quark scaling of $v_{2}$, it is clear that
$\rho^{0}$-vector meson follows $n=2$ quarks in the intermediate
$p_{T}$ range which implies most of the $\rho^{0}$s are formed from
the quarks coalescence.

\end{document}